# ANÁLISIS DE DISTANCIAS TEMPORALES Y ESPACIALES ENTRE EL LUGAR DE LA MANCHA Y CUATRO PUNTOS DE REFERENCIA

Por D. Orden Martín y R. Viaña Fernández.
david.orden@uah.es, raquel.viana@uah.es
Departamento de Matemáticas, Universidad de Alcalá

## 1. Introducción

La identidad del famoso lugar de La Mancha que se menciona en la primera frase de *El ingenioso hidalgo Don Quijote de La Mancha* es una incógnita con una historia casi tan larga como la de la famosa obra de Miguel de Cervantes. En particular, en los últimos años han aparecido diversos trabajos con el objetivo de resolver esta cuestión, la mayoría de ellos a partir de datos sobre la ubicación del lugar de La Mancha similares a los siguientes que, según las conclusiones de [5] refinadas posteriormente en [6], aparecen en la novela cervantina:

- Está situado en la comarca histórica denominada Campo de Montiel. En concreto, es una de las 24 localidades que aparecen en la Tabla 1.
- Está a 2 jornadas de la Venta de Cárdenas.
- Está a 2.37 jornadas de Puerto Lápice.
- Está a 2.5 jornadas de El Toboso.
- Está a 2 jornadas de Munera.

Además, también según [5], la velocidad de los protagonistas de la novela está comprendida entre 30 y 35 kilómetros por jornada.

Permítasenos observar aquí que, a pesar de que en [6] aparece un segundo conjunto de datos hipótesis sobre la ubicación del lugar de La Mancha, con el fin de no entorpecer el hilo argumental comenzaremos trabajando sobre los datos anteriormente mostrados, que son los más similares a los utilizados en los trabajos previos. Será en una sección posterior cuando se muestre el nuevo conjunto de datos y se realice su análisis.

Si se aceptan estos datos como ciertos (la discusión de este punto está fuera del objetivo del presente trabajo), el método más natural de resolver el problema se puede explicar fácilmente mediante la siguiente analogía. Imaginemos un sorteo de lotería en el que cada una de las 24 localidades del Campo de Montiel enumeradas en la Tabla 1 posee un boleto con una combinación de cuatro números, y que éstos son el número de jornadas que separan dicha localidad de cada uno de los cuatro puntos de referencia anteriormente citados, Venta de Cárdenas, Puerto Lápice, El Toboso y Munera. Según los datos arriba expuestos, la localidad agraciada con el hecho de ser el lugar de La Mancha será la propietaria del boleto con la combinación (2,2.37,2.5,2).

Sin embargo, ¿qué sucede si ninguna de esas 24 localidades posee esa combinación? En ese caso, uno puede plantearse cuál de ellas tiene la combinación que más se aproxima a la combinación premiada. Para ello hay varios criterios, todos ellos igual de correctos, en función de qué se entienda por "aproximarse más". Estos criterios se corresponden con la noción matemática de *métrica*.

## 2. Métricas $L_\infty$, $L_1$, $L_2$, y $L_n$

Si el lector tuviera un boleto con la combinación (a,b,c,d) y supiera que la combinación premiada es (2,2.37,2.5,2), ¿cómo mediría cuán cerca está su combinación de la ganadora? A continuación veremos diferentes opciones; para comprender mejor cada una de ellas, se aconseja que el lector escoja sus propios valores de (a,b,c,d) y realice los cálculos propuestos.

- **Métrica $L_\infty$:** Una primera manera de medir sería comprobar cuál de las cuatro casillas de nuestra combinación discrepa más de la correspondiente casilla de la combinación ganadora, y tomar esa discrepancia como medida de aproximación a la combinación premiada. Así, la distancia $L_\infty$ entre (a,b,c,d) y (2,2.37,2.5,2), que notaremos $d_\infty((a,b,c,d),(2,2.37,2.5,2))$, se define como

$$d_\infty((a,b,c,d),(2,2.37,2.5,2)) = max\{|a-2|,|b-2.37|,|c-2.5|,|d-2|\}$$

- **Métrica $L_1$:** La siguiente manera de medir que vamos a proponer consiste en sumar las discrepancias entre cada una de las cuatro casillas de nuestra combinación y la correspondiente casilla de la combinación ganadora. Se define la distancia $L_1$ entre (a,b,c,d) y (2,2.37,2.5,2), que notaremos $d_1((a,b,c,d),(2,2.37,2.5,2))$, como

$$d_1((a,b,c,d),(2,2.37,2.5,2)) = |a-2|+|b-2.37|+|c-2.5|+|d-2|$$

- **Métrica $L_2$:** Esta forma de medir consiste en realizar la raíz cuadrada de la suma de los cuadrados de las discrepancias entre cada una de las cuatro casillas de nuestra combinación y la correspondiente casilla de la combinación ganadora. De esta manera, se define la distancia $L_2$ o distancia euclídea $d_2$ entre (a,b,c,d) y (2,2.37, 2.5,2) como

$$d_2((a,b,c,d),(2,2.37,2.5,2)) = \sqrt{|a-2|^2 + |b-2.37|^2 + |c-2.5|^2 + |d-2|^2}$$

- **Métrica $L_n$:** Las métricas $L_1$ y $L_2$ son en realidad casos particulares de la métrica $L_n$, definida para cualquier número natural $n$. La distancia $L_n$ entre (a,b,c,d) y (2,2.37, 2.5,2), que notaremos $d_n((a,b,c,d),(2,2.37,2.5,2))$, está definida como

$$d_n((a,b,c,d),(2,2.37,2.5,2)) = (|a-2|^n+|b-2.37|^n+|c-2.5|^n+|d-2|^n)^{1/n}$$

## 3. Trabajos previos

El método de resolución seguido en [7] es similar al expuesto en las dos secciones anteriores. La diferencia principal es que, en lugar de trabajar con distancias en términos de número de jornadas, se trabaja con distancias kilométricas. Por una parte, para cada una de las 24 localidades del Campo de Montiel su combinación (a,b,c,d) contiene la distancia en kilómetros a cada uno de los cuatro puntos de referencia, medida en línea recta y sobre el plano según el mapa de carreteras Michelin. Por otra parte, la combinación solución (2,2,2.5,2) que consideran, expresada en jornadas, debe traducirse a kilómetros para poder compararla con las anteriores. Para ello se consideran valores de la velocidad $v$ de los protagonistas entre 25 y 40 kilómetros por jornada y se estudia cuál de los valores de *v* considerados hace que las discrepancias con respecto a la solución sean mínimas. Para la métrica $L_1$ resulta una velocidad de 34 kms/jornada, y la localidad cuya combinación está más cerca de la solución resulta ser Alcubillas, con una discrepancia de 24 unidades, mientras que Villanueva de los Infantes aparece en segundo lugar con una discrepancia de 30 unidades. Para la métrica $L_2$ resulta una velocidad de 34.37 kms/jornada, obteniendo de nuevo como localidad candidata Alcubillas. Finalmente, para la métrica $L_\infty$ obtienen una velocidad de 32.75 kms/jornada, que conduce a Villanueva de los Infantes como localidad candidata.

En [2] parten de la combinación solución (2,2.25,2.5,1.75). Se aplican técnicas estadísticas a estas distancias en jornadas y a las distancias kilométricas entre los cuatro puntos de referencia y las localidades del Campo de Montiel, medidas también en línea recta sobre el mapa de carreteras Michelin. Con ello se obtiene una estimación de las distancias espaciales desde el lugar de La Mancha a cada uno de los cuatro puntos de referencia, que se utilizan para ordenar las 24 localidades del Campo de Montiel según su probabilidad de ser el lugar de La Mancha. El resultado establece Villanueva de los Infantes como la localidad más probable, con una probabilidad de 0.15992 (sobre 1), mientras que la segunda localidad más probable es Fuenllana, con una probabilidad de 0.15433.

En [1] se consideran de nuevo las distancias kilométricas a los cuatro puntos de referencia desde las localidades del Campo de Montiel, pero esta vez se utiliza como fuente el Directorio Cartográfico de España. Toman la siguiente combinación solución en horas, con una confianza del 95%: (20±15%,22.5±15%,25±10%,20±10%). Con estos datos se aplican técnicas de Teoría de Sistemas para tratar de establecer tanto el lugar de La Mancha como la velocidad de los protagonistas. Con una fiabilidad del 95%, Villanueva de los Infantes resulta ser la localidad más probable, con probabilidad entre 0.353 y 0.358 para una velocidad entre 2.98 y 3.11 kilómetros por hora. La segunda localidad más probable es Alhambra, con probabilidad entre 0.31 y 0.31 para una velocidad entre 2.69 y 2.82 kilómetros por hora.

En [3] se vuelven a considerar las distancias kilométricas con el Directorio Cartográfico de España como fuente. Con ellas, para cada localidad del Campo de Montiel se calcula su discrepancia con respecto a las distancias para el lugar de La Mancha, obtenidas a partir de las tardanzas en jornadas dadas en [5] y la velocidad media *v* de 31 kilómetros por jornada, lo que dada la dificultad de aproximar la distancia del lugar de La Mancha a Puerto Lápice conduce a una de las siguientes combinaciones solución en kilómetros: (62,69.4,77.5,61.7), (62,70,77.5,61.7), (62,71,77.5,61.7), (62,72,77.5,61.7), (62,73,77.5,61.7) ó (62,73.5,77.5,61.7). Para medir esta discrepancia se utilizan las métricas $L_1$, $L_2$ y $L_\infty$, obteniendo los mismos resultados para todas las combinaciones solución consideradas. Para la métrica $L_1$, la menor discrepancia se da para Carrizosa, con 18.55 unidades, mientras que Fuenllana tiene 24.96 y Villanueva de los Infantes tiene 25.04. Para la métrica $L_2$ se obtiene Carrizosa con 12.44, Villanueva de los Infantes con 13.46 y Fuenllana con 13.56. Finalmente, la métrica $L_\infty$ arroja como resultados para estas tres localidades Fuenllana con 9.3, Villanueva de los Infantes con 9.44 y Carrizosa con 11.72.

Por último en [4] parten de la combinación solución (2,2.37,2.5,2) y de las distancias entre las cuatro referencias y el lugar de La Mancha estimadas a partir del Directorio Cartográfico de España. Se analizan, además de las distancias, velocidades medias plausibles, obteniendo como resultado Villanueva de los Infantes con 30.59 kilómetros por jornada y Fuenllana con 30.57 kilómetros por jornada.

## 4. Aportación y organización de este trabajo

Como se ha mostrado en la sección anterior, los trabajos realizados hasta el momento coinciden en utilizar distancias kilométricas entre las 24 localidades del Campo de Montiel y las cuatro localidades de referencia. Siguiendo esta línea, en el trabajo [8] del presente volumen se ha utilizado un Sistema de Información Geográfica para, teniendo en cuenta accidentes geográficos de los caminos, establecer la distancia entre cada una de las 24 localidades del Campo de Montiel y las cuatro referencias mediante caminos óptimos, tanto en kilómetros como en jornadas.

En el presente trabajo dejaremos a un lado la discusión sobre la corrección de los datos aportados en [5], [6] y [8], que se deja para expertos en otras disciplinas. Aquí nos limitaremos a realizar un análisis como el introducido en las Secciones 1 y 2 para los datos obtenidos en [8], sin más pretensiones que utilizar herramientas matemáticas básicas para determinar cuál de las 24 localidades del Campo de Montiel posee la combinación de cuatro distancias más cercana a una combinación dada como solución, que corresponde a las cuatro referencias Puerto Lápice, Venta de Cárdenas, El Toboso y Munera. Además, a petición de los autores de [6] se considerará también el caso con sólo tres puntos de referencia en el que se ha descartado la correspondiente a Munera.

Así, por un lado utilizaremos los datos sobre distancias en jornadas y en kilómetros obtenidos en [8]. Por otro lado, consideraremos las dos combinaciones solución propuestas en [5] y [6], medidas en jornadas, junto con su conversión a kilómetros. Las tardanzas en jornadas serán la ya mencionada (2,2.37,2.5,2), comúnmente utilizada hasta el momento y la (2,2.42,2.8,2.23), propuesta en el presente volumen [6] y que contabiliza de manera más ajustada el número de horas de marcha durante la noche. La conversión a horas se realizará utilizando la duración de 10 horas por jornada [6]. La conversión a kilómetros se realizará utilizando la velocidad media de 31 kilómetros por jornada establecida también en [6]. En cuanto a las métricas consideradas, el hecho de que exista una métrica $L_n$ para cada número natural $n$ hace imposible considerar todas ellas, por lo que se utilizarán las métricas $L_\infty$, $L_1$, y $L_2$. Finalmente, con el fin de facilitar la lectura, en cada caso se mostrarán únicamente los resultados para las cinco localidades más cercanas a la solución.

La organización del resto de este trabajo es la siguiente: La Sección 5 está dedicada a la combinación solución considerada en los trabajos anteriores. Las Subsecciones 5.1 y 5.2 contienen el estudio para las tardanzas en jornadas y en kilómetros, respectivamente, considerando las cuatro localidades de referencia ya mencionadas. Las Subsecciones 5.3 y 5.4 realizan el estudio análogo considerando las tres localidades de referencia resultantes de descartar Munera. La Sección 6 se estructura de manera análoga, pero está dedicada a la nueva combinación solución que se introduce en este volumen [6]. La Sección 7 compara los resultados de este estudio con los de trabajos anteriores y, finalmente, la Sección 8 contiene las conclusiones obtenidas.

## 5. Análisis para tardanzas en jornadas (2,2.37,2.5,2)

La presente sección utiliza las tardanzas ya utilizadas en la mayoría de los trabajos anteriores. En la Subsección 5.1 trabajaremos con las distancias kilométricas desde cada una de las cuatro referencias a las 24 localidades del Campo de Montiel que se encuentran recogidas en la Tabla 1. A continuación, en la Subsección 5.2, realizaremos un análisis similar pero partiendo de los datos temporales de la Tabla 5. Finalmente, en las dos últimas subsecciones repetiremos el proceso eliminando Munera, es decir, tomando solo Puerto Lápice, Venta de Cárdenas y El Toboso como localidades de referencia.

### 5.1. Análisis espacial con cuatro referencias

En la Tabla 1 se recogen las distancias kilométricas que forman los datos de partida. De este modo cada localidad posee una combinación de cuatro distancias, que deberemos comparar con la combinación solución. En esta subsección se considera la combinación solución en jornadas (2,2.37,2.5,2), que convertida a kilómetros a razón de 31 kilómetros por jornada [6] resulta (62,73.47,77.5,62).

|  | Venta de Cárdenas | Puerto Lápice | El Toboso | Munera |
|---|---|---|---|---|
|  | **km** | **km** | **km** | **km** |
| Albaladejo | **72,80** | **94,40** | **106,92** | **53,68** |
| Alcubillas | **55,88** | **66,76** | **86,64** | **67,08** |
| Alhambra | **72,08** | **64,04** | **68,72** | **53,16** |
| Almedina | **59,28** | **84,04** | **99,40** | **65,04** |
| Cañamares | **78,96** | **94,48** | **102,16** | **46,16** |
| Carrizosa | **70,44** | **72,28** | **77,20** | **52,52** |
| Castellar de Santiago | **30,00** | **94,48** | **116,52** | **92,28** |
| Cózar | **57,52** | **77,72** | **95,72** | **67,44** |
| Fuenllana | **71,56** | **76,36** | **87,00** | **55,68** |
| Membrilla | **74,88** | **39,44** | **76,00** | **78,44** |
| Montiel | **68,36** | **84,92** | **97,44** | **56,72** |
| Ossa de Montiel | **98,68** | **75,08** | **68,68** | **23,60** |
| Puebla del Príncipe | **61,44** | **90,80** | **106,16** | **65,04** |
| Ruidera | **86,92** | **65,96** | **64,64** | **36,04** |
| Sta. Cruz de Cáñamos | **66,96** | **90,84** | **104,40** | **57,28** |
| La Solana | **70,44** | **47,84** | **66,76** | **69,20** |
| Terrinches | **69,64** | **94,16** | **107,72** | **56,84** |
| Torre de Juan Abad | **49,12** | **86,04** | **104,16** | **73,16** |
| Torres de Montiel | **66,36** | **80,32** | **95,68** | **90,32** |
| Torrenueva | **32,64** | **81,68** | **103,72** | **59,04** |
| Villahermosa | **74,00** | **83,76** | **91,44** | **48,28** |
| Villamanrique | **55,12** | **92,08** | **108,20** | **71,56** |
| Villanueva de la Fuente | **82,00** | **99,48** | **107,48** | **42,16** |
| Villanueva de los Infantes | **66,24** | **71,48** | **87,04** | **61,00** |

Tabla 1. Distancias en kilómetros, siguiendo los caminos óptimos dados en [8], entre las cuatro referencias y las 24 localidades candidatas.

A continuación se muestran los resultados de estas comparaciones utilizando las métricas explicadas en la Sección 2. Así, en las Tablas 2, 3 y 4 se muestran los resultados de estas comparaciones utilizando las métricas $L_\infty$, $L_1$, y $L_2$, respectivamente.

|  | Venta de Cárdenas | Puerto Lápice | El Toboso | Munera | $d_\infty$ |
|---|---|---|---|---|---|
|  | **km** | **km** | **km** | **km** |  |
| LUGAR DE LA MANCHA | 62,00 | 73,47 | 77,50 | 62,00 | 0,00 |
| Alcubillas | 55,88 | 66,76 | 86,64 | 67,08 | 9,14 |
| Carrizosa | 70,44 | 72,28 | 77,20 | 52,52 | 9,48 |
| Villanueva de los Infantes | 66,24 | 71,48 | 87,04 | 61,00 | 9,54 |
| Fuenllana | 71,56 | 76,36 | 87,00 | 55,68 | 9,56 |
| Alhambra | 72,08 | 64,04 | 68,72 | 53,16 | 10,08 |

Tabla 2. Distancias $d_\infty$ a la combinación en kilómetros correspondiente a (2,2.37,2.5,2) jornadas, para las cinco localidades con combinación más cercana.

|  | Venta de Cárdenas | Puerto Lápice | El Toboso | Munera | $d_1$ |
|---|---|---|---|---|---|
|  | **km** | **km** | **km** | **km** |  |
| LUGAR DE LA MANCHA | 62,00 | 73,47 | 77,50 | 62,00 | 0,00 |
| Villanueva de los Infantes | 66,24 | 71,48 | 87,04 | 61,00 | 16,77 |
| Carrizosa | 70,44 | 72,28 | 77,20 | 52,52 | 19,41 |
| Alcubillas | 55,88 | 66,76 | 86,64 | 67,08 | 27,05 |
| Fuenllana | 71,56 | 76,36 | 87,00 | 55,68 | 28,27 |
| Cózar | 57,52 | 77,72 | 95,72 | 67,44 | 32,39 |

Tabla 3. Distancias $d_1$ a la combinación en kilómetros correspondiente a (2,2.37,2.5,2) jornadas, para las cinco localidades con combinación más cercana.

|  | Venta de Cárdenas | Puerto Lápice | El Toboso | Munera | $d_2$ |
|---|---|---|---|---|---|
|  | **km** | **km** | **km** | **km** |  |
| LUGAR DE LA MANCHA | 62,00 | 73,47 | 77,50 | 62,00 | 0,00 |
| Villanueva de los Infantes | 66,24 | 71,48 | 87,04 | 61,00 | 10,67 |
| Carrizosa | 70,44 | 72,28 | 77,20 | 52,52 | 12,75 |
| Alcubillas | 55,88 | 66,76 | 86,64 | 67,08 | 13,85 |
| Fuenllana | 71,56 | 76,36 | 87,00 | 55,68 | 15,16 |
| Alhambra | 72,08 | 64,04 | 68,72 | 53,16 | 18,59 |

Tabla 4. Distancias $d_2$ a la combinación en kilómetros correspondiente a (2,2.37,2.5,2) jornadas, para las cinco localidades con combinación más cercana.

### 5.2. Análisis temporal con cuatro referencias

En esta subsección se consideran las distancias temporales, en horas, recogidas en la Tabla 5, y se analiza su discrepancia con la combinación solución en jornadas (2,2.37,2.5,2), convertida a horas a razón de 10 horas por jornada [6].

|  | Venta de Cárdenas | Puerto Lápice | El Toboso | Munera |
|---|---|---|---|---|
|  | **horas** | **horas** | **horas** | **horas** |
| Albaladejo | 23,48 | 30,45 | 34,49 | 17,32 |
| Alcubillas | 18,03 | 21,54 | 27,95 | 21,64 |
| Alhambra | 23,25 | 20,66 | 22,17 | 17,15 |
| Almedina | 19,12 | 27,11 | 32,06 | 20,98 |
| Cañamares | 25,47 | 30,48 | 32,95 | 14,89 |
| Carrizosa | 22,72 | 23,32 | 24,90 | 16,94 |
| Castellar de Santiago | 9,68 | 30,48 | 37,59 | 29,77 |
| Cózar | 18,55 | 25,07 | 30,88 | 21,75 |
| Fuenllana | 23,08 | 24,63 | 28,06 | 17,96 |
| Membrilla | 24,15 | 12,72 | 24,52 | 25,30 |
| Montiel | 22,05 | 27,39 | 31,43 | 18,30 |
| Ossa de Montiel | 31,83 | 24,22 | 22,15 | 7,61 |
| Puebla del Príncipe | 19,82 | 29,29 | 34,25 | 20,98 |
| Ruidera | 28,04 | 21,28 | 20,85 | 11,63 |
| Sta. Cruz de Cáñamos | 21,60 | 29,30 | 33,68 | 18,48 |
| La Solana | 22,72 | 15,43 | 21,54 | 22,32 |
| Terrinches | 22,46 | 30,37 | 34,75 | 18,34 |
| Torre de Juan Abad | 15,85 | 27,75 | 33,60 | 23,60 |
| Torres de Montiel | 21,41 | 25,91 | 30,86 | 29,14 |
| Torrenueva | 10,53 | 26,35 | 33,46 | 19,05 |
| Villahermosa | 23,87 | 27,02 | 29,50 | 15,57 |
| Villamanrique | 17,78 | 29,70 | 34,90 | 23,08 |
| Villanueva de la Fuente | 26,45 | 32,09 | 34,67 | 13,60 |
| Villanueva de los Infantes | 21,37 | 23,06 | 28,08 | 19,68 |

Tabla 5. Distancia en horas entre cada una de las 24 localidades del Campo de Montiel y los cuatro puntos de referencia, según [8].

A continuación, en las Tablas 6, 7 y 8 se muestran los resultados de estas comparaciones utilizando las métricas $L_\infty$, $L_1$ y $L_2$ respectivamente.

|  | Venta de Cárdenas | Puerto Lápice | El Toboso | Munera | $d_\infty$ |
|---|---|---|---|---|---|
|  | **horas** | **horas** | **horas** | **horas** |  |
| LUGAR DE LA MANCHA | 20,00 | 23,70 | 25,00 | 20,00 | 0,00 |
| Alcubillas | 18,03 | 21,54 | 27,95 | 21,64 | 2,95 |
| Carrizosa | 22,72 | 23,32 | 24,90 | 16,94 | 3,06 |
| Villanueva de los Infantes | 21,37 | 23,06 | 28,08 | 19,68 | 3,08 |
| Fuenllana | 23,08 | 24,63 | 28,06 | 17,96 | 3,08 |
| Alhambra | 23,25 | 20,66 | 22,17 | 17,15 | 3,25 |

Tabla 6. Distancias $d_\infty$ a la combinación en horas correspondiente a (2,2.37,2.5,2) jornadas, para las cinco localidades con combinación más cercana.

|  | Venta de Cárdenas | Puerto Lápice | El Toboso | Munera | $d_1$ |
|---|---|---|---|---|---|
|  | **horas** | **horas** | **horas** | **horas** |  |
| LUGAR DE LA MANCHA | **20,00** | **23,70** | **25,00** | **20,00** | **0,00** |
| Villanueva de los Infantes | **21,37** | **23,06** | **28,08** | **19,68** | 5,41 |
| Carrizosa | **22,72** | **23,32** | **24,90** | **16,94** | 6,26 |
| Alcubillas | **18,03** | **21,54** | **27,95** | **21,64** | 8,72 |
| Fuenllana | **23,08** | **24,63** | **28,06** | **17,96** | 9,11 |
| Cózar | **18,55** | **25,07** | **30,88** | **21,75** | 10,45 |

Tabla 7. Distancias $d_1$ a la combinación en horas correspondiente a (2,2.37,2.5,2) jornadas, para las cinco localidades con combinación más cercana.

|  | Venta de Cárdenas | Puerto Lápice | El Toboso | Munera | $d_2$ |
|---|---|---|---|---|---|
|  | **horas** | **horas** | **horas** | **horas** |  |
| LUGAR DE LA MANCHA | **20,00** | **23,70** | **25,00** | **20,00** | **0,00** |
| Villanueva de los Infantes | **21,37** | **23,06** | **28,08** | **19,68** | 3,45 |
| Carrizosa | **22,72** | **23,32** | **24,90** | **16,94** | 4,11 |
| Alcubillas | **18,03** | **21,54** | **27,95** | **21,64** | 4,47 |
| Fuenllana | **23,08** | **24,63** | **28,06** | **17,96** | 4,89 |
| Alhambra | **23,25** | **20,66** | **22,17** | **17,15** | 5,99 |

Tabla 8. Distancias $d_2$ a la combinación en horas correspondiente a (2,2.37,2.5,2) jornadas, para las cinco localidades con combinación más cercana.

### 5.3. Análisis espacial con 3 referencias

En esta subsección se realiza un análisis análogo al de la Subsección 5.1 pero excluyendo la columna correspondiente a Munera. De este modo, cada localidad posee una combinación de tres distancias, que deberemos comparar con la combinación solución. Presentamos a continuación, en las Tablas 9, 10 y 11, los resultados de estas comparaciones utilizando las métricas $L_\infty$, $L_1$, y $L_2$, respectivamente.

|  | Venta de Cárdenas | Puerto Lápice | El Toboso | $d_\infty$ |
|---|---|---|---|---|
|  | **km** | **km** | **km** |  |
| LUGAR DE LA MANCHA | **62,00** | **73,47** | **77,50** | **0,00** |
| Carrizosa | **70,44** | **72,28** | **77,20** | 8,44 |
| Alcubillas | **55,88** | **66,76** | **86,64** | 9,14 |
| Villanueva de los Infantes | **66,24** | **71,48** | **87,04** | 9,54 |
| Fuenllana | **71,56** | **76,36** | **87,00** | 9,56 |
| Alhambra | **72,08** | **64,04** | **68,72** | 10,08 |

Tabla 9. Distancias $d_\infty$ a la combinación en kilómetros correspondiente a (2,2.37,2.5) jornadas, para las cinco localidades con combinación más cercana.

|  | Venta de Cárdenas | Puerto Lápice | El Toboso | $d_1$ |
|---|---|---|---|---|
|  | **km** | **km** | **km** |  |
| LUGAR DE LA MANCHA | **62,00** | **73,47** | **77,50** | **0,00** |
| Carrizosa | **70,44** | **72,28** | **77,20** | 9,93 |
| Villanueva de los Infantes | **66,24** | **71,48** | **87,04** | 15,77 |
| Fuenllana | **71,56** | **76,36** | **87,00** | 21,95 |
| Alcubillas | **55,88** | **66,76** | **86,64** | 21,97 |
| Cózar | **57,52** | **77,72** | **95,72** | 26,95 |

Tabla 10. Distancias $d_1$ a la combinación en kilómetros correspondiente a (2,2.37,2.5) jornadas, para las cinco localidades con combinación más cercana.

|  | Venta de Cárdenas | Puerto Lápice | El Toboso | $d_2$ |
|---|---|---|---|---|
|  | **km** | **km** | **km** |  |
| LUGAR DE LA MANCHA | **62,00** | **73,47** | **77,50** | **0,00** |
| Carrizosa | **70,44** | **72,28** | **77,20** | 8,53 |
| Villanueva de los Infantes | **66,24** | **71,48** | **87,04** | 10,63 |
| Alcubillas | **55,88** | **66,76** | **86,64** | 12,88 |
| Fuenllana | **71,56** | **76,36** | **87,00** | 13,78 |
| Alhambra | **72,08** | **64,04** | **68,72** | 16,36 |

Tabla 11. Distancias $d_2$ a la combinación en kilómetros correspondiente a (2,2.37,2.5) jornadas, para las cinco localidades con combinación más cercana.

### 5.4. Análisis temporal con 3 referencias.

En esta subsección, siguiendo el mismo espíritu de la anterior, se realiza un análisis análogo al de la Subsección 5.2 pero eliminando los datos correspondientes a Munera. Las Tablas 12, 13 y 14 muestran los resultados de las comparaciones utilizando las métricas $L_\infty$, $L_1$ y $L_2$ respectivamente.

|  | Venta de Cárdenas | Puerto Lápice | El Toboso | $d_\infty$ |
|---|---|---|---|---|
|  | **horas** | **horas** | **horas** |  |
| LUGAR DE LA MANCHA | **20,00** | **23,70** | **25,00** | **0,00** |
| Carrizosa | **22,72** | **23,32** | **24,90** | 2,72 |
| Alcubillas | **18,03** | **21,54** | **27,95** | 2,95 |
| Villanueva de los Infantes | **21,37** | **23,06** | **28,08** | 3,08 |
| Fuenllana | **23,08** | **24,63** | **28,06** | 3,08 |
| Alhambra | **23,25** | **20,66** | **22,17** | 3,25 |

Tabla 12. Distancias $d_\infty$ a la combinación en horas correspondiente a (2,2.37,2.5) jornadas, para las cinco localidades con combinación más cercana.

|  | Venta de Cárdenas | Puerto Lápice | El Toboso | $d_1$ |
|---|---|---|---|---|
|  | **horas** | **horas** | **horas** |  |
| LUGAR DE LA MANCHA | **20,00** | **23,70** | **25,00** | **0,00** |
| Carrizosa | **22,72** | **23,32** | **24,90** | 3,20 |
| Villanueva de los Infantes | **21,37** | **23,06** | **28,08** | 5,09 |
| Fuenllana | **23,08** | **24,63** | **28,06** | 7,07 |
| Alcubillas | **18,03** | **21,54** | **27,95** | 7,08 |
| Cózar | **18,55** | **25,07** | **30,88** | 8,70 |

Tabla 13. Distancias $d_1$ a la combinación en horas correspondiente a (2,2.37,2.5) jornadas, para las cinco localidades con combinación más cercana.

|  | Venta de Cárdenas | Puerto Lápice | El Toboso | $d_2$ |
|---|---|---|---|---|
|  | **horas** | **horas** | **horas** |  |
| LUGAR DE LA MANCHA | **20,00** | **23,70** | **25,00** | **0,00** |
| Carrizosa | **22,72** | **23,32** | **24,90** | 2,75 |
| Villanueva de los Infantes | **21,37** | **23,06** | **28,08** | 3,43 |
| Alcubillas | **18,03** | **21,54** | **27,95** | 4,15 |
| Fuenllana | **23,08** | **24,63** | **28,06** | 4,44 |
| Alhambra | **23,25** | **20,66** | **22,17** | 5,27 |

Tabla 14. Distancias $d_2$ a la combinación en horas correspondiente a (2,2.37,2.5) jornadas, para las cinco localidades con combinación más cercana.

## 6. Análisis para tardanzas en jornadas (2,2.42,2.8,2.23)

En esta sección pasamos a considerar como combinación solución la introducida en el presente volumen [6]. El esquema es el mismo de la Sección 5. Así, la Subsección 6.1 considera las distancias kilométricas desde cada una de las cuatro referencias a las 24 localidades del Campo de Montiel recogidas en la Tabla 1. A continuación, en la Subsección 6.2, realizaremos un análisis similar pero partiendo de los datos temporales que se encuentran en la Tabla 5. Finalmente, en las dos últimas subsecciones repetiremos el proceso eliminando Munera, es decir, tomando sólo Puerto Lápice, Venta de Cárdenas y El Toboso como localidades de referencia.

### 6.1. Análisis espacial con 4 referencias

Recordemos en este punto las distancias kilométricas recogidas en la Tabla 1, de modo que cada localidad posee una combinación de cuatro distancias, que compararemos con la combinación solución. En esta subsección se considera la combinación solución en jornadas (2,2.42,2.8,2.23), que convertida a kilómetros a razón de 31 kilómetros por jornada [6] resulta (62,75.02,86.80,69.13). Las Tablas 15, 16 y 17 muestran los resultados de estas comparaciones utilizando las métricas $L_\infty$, $L_1$ y $L_2$ respectivamente.

|  | Venta de Cárdenas | Puerto Lápice | El Toboso | Munera | $d_\infty$ |
|---|---|---|---|---|---|
|  | km | km | km | km |  |
| LUGAR DE LA MANCHA | 62,00 | 75,02 | 86,80 | 69,13 | 0,00 |
| Villanueva de los Infantes | 66,24 | 71,48 | 87,04 | 61,00 | 8,13 |
| Alcubillas | 55,88 | 66,76 | 86,64 | 67,08 | 8,26 |
| Cózar | 57,52 | 77,72 | 95,72 | 67,44 | 8,92 |
| Montiel | 68,36 | 84,92 | 97,44 | 56,72 | 12,41 |
| Almedina | 59,28 | 84,04 | 99,40 | 65,04 | 12,60 |

Tabla 15. Distancias $d_\infty$ a la combinación en kilómetros correspondiente a (2,2.42,2.8,2.23) jornadas, para las cinco localidades con combinación más cercana.

|  | Venta de Cárdenas | Puerto Lápice | El Toboso | Munera | $d_1$ |
|---|---|---|---|---|---|
|  | km | km | km | km |  |
| LUGAR DE LA MANCHA | 62,00 | 75,02 | 86,80 | 69,13 | 0,00 |
| Villanueva de los Infantes | 66,24 | 71,48 | 87,04 | 61,00 | 16,15 |
| Alcubillas | 55,88 | 66,76 | 86,64 | 67,08 | 16,59 |
| Cózar | 57,52 | 77,72 | 95,72 | 67,44 | 17,79 |
| Fuenllana | 71,56 | 76,36 | 87,00 | 55,68 | 24,55 |
| Almedina | 59,28 | 84,04 | 99,40 | 65,04 | 28,43 |

Tabla 16. Distancias $d_1$ a la combinación en kilómetros correspondiente a (2,2.42,2.8,2.23) jornadas, para las cinco localidades con combinación más cercana.

|  | Venta de Cárdenas | Puerto Lápice | El Toboso | Munera | $d_2$ |
|---|---|---|---|---|---|
|  | km | km | km | km |  |
| LUGAR DE LA MANCHA | 62,00 | 75,02 | 86,80 | 69,13 | 0,00 |
| Villanueva de los Infantes | 66,24 | 71,48 | 87,04 | 61,00 | 9,83 |
| Cózar | 57,52 | 77,72 | 95,72 | 67,44 | 10,48 |
| Alcubillas | 55,88 | 66,76 | 86,64 | 67,08 | 10,48 |
| Almedina | 59,28 | 84,04 | 99,40 | 65,04 | 16,26 |
| Fuenllana | 71,56 | 76,36 | 87,00 | 55,68 | 16,56 |

Tabla 17. Distancias $d_2$ a la combinación en kilómetros correspondiente a (2,2.42,2.8,2.23) jornadas, para las cinco localidades con combinación más cercana.

## 6.2. Análisis temporal con 4 referencias

En esta subsección se consideran las distancias temporales, en horas, recogidas en la Tabla 5, y se analiza su discrepancia con la combinación solución en jornadas (2,2.42,2.8,2.23), convertida a horas a razón de 10 horas por jornada. Las Tablas 18, 19 y 20 muestran los resultados de estas comparaciones utilizando las métricas $L_\infty$, $L_1$ y $L_2$ respectivamente.

|                           | Venta de Cárdenas | Puerto Lápice | El Toboso | Munera | $d_\infty$ |
|---------------------------|-------------------|---------------|-----------|--------|------------|
|                           | **horas**         | **horas**     | **horas** | **horas** |         |
| LUGAR DE LA MANCHA        | **20,00**         | **24,20**     | **28,00** | **22,30** | **0,00** |
| Villanueva de los Infantes| **21,37**         | **23,06**     | **28,08** | **19,68** | 2,62 |
| Alcubillas                | **18,03**         | **21,54**     | **27,95** | **21,64** | 2,66 |
| Cózar                     | **18,55**         | **25,07**     | **30,88** | **21,75** | 2,88 |
| Montiel                   | **22,05**         | **27,39**     | **31,43** | **18,30** | 4,00 |
| Almedina                  | **19,12**         | **27,11**     | **32,06** | **20,98** | 4,06 |

Tabla 18. Distancias $d_\infty$ a la combinación en horas correspondiente a (2,2.42,2.8,2.23) jornadas, para las cinco localidades con combinación más cercana.

|                           | Venta de Cárdenas | Puerto Lápice | El Toboso | Munera | $d_1$ |
|---------------------------|-------------------|---------------|-----------|--------|-------|
|                           | **horas**         | **horas**     | **horas** | **horas** |    |
| LUGAR DE LA MANCHA        | **20,00**         | **24,20**     | **28,00** | **22,30** | **0,00** |
| Villanueva de los Infantes| **21,37**         | **23,06**     | **28,08** | **19,68** | 5,21 |
| Alcubillas                | **18,03**         | **21,54**     | **27,95** | **21,64** | 5,34 |
| Cózar                     | **18,55**         | **25,07**     | **30,88** | **21,75** | 5,75 |
| Fuenllana                 | **23,08**         | **24,63**     | **28,06** | **17,96** | 7,91 |
| Almedina                  | **19,12**         | **27,11**     | **32,06** | **20,98** | 9,17 |

Tabla 19. Distancias $d_1$ a la combinación en horas correspondiente a (2,2.42,2.8,2.23) jornadas, para las cinco localidades con combinación más cercana.

|                           | Venta de Cárdenas | Puerto Lápice | El Toboso | Munera | $d_2$ |
|---------------------------|-------------------|---------------|-----------|--------|-------|
|                           | **horas**         | **horas**     | **horas** | **horas** |    |
| LUGAR DE LA MANCHA        | **20,00**         | **24,20**     | **28,00** | **22,30** | **0,00** |
| Villanueva de los Infantes| **21,37**         | **23,06**     | **28,08** | **19,68** | 3,17 |
| Alcubillas                | **18,03**         | **21,54**     | **27,95** | **21,64** | 3,38 |
| Cózar                     | **18,55**         | **25,07**     | **30,88** | **21,75** | 3,38 |
| Almedina                  | **19,12**         | **27,11**     | **32,06** | **20,98** | 5,24 |
| Fuenllana                 | **23,08**         | **24,63**     | **28,06** | **17,96** | 5,34 |

Tabla 20. Distancias $d_2$ a la combinación en horas correspondiente a (2,2.42,2.8,2.23) jornadas, para las cinco localidades con combinación más cercana.

### 6.3. Análisis espacial con 3 referencias

En esta subsección se realiza un estudio análogo al de la Subsección 6.1, pero excluyendo Munera como punto de referencia. En las Tablas 21, 22 y 23 se muestran los resultados de estas comparaciones utilizando las métricas $L_\infty$, $L_1$ y $L_2$ respectivamente.

|                            | Venta de Cárdenas | Puerto Lápice | El Toboso | $d_\infty$ |
|----------------------------|-------------------|---------------|-----------|------------|
|                            | **km**            | **km**        | **km**    |            |
| LUGAR DE LA MANCHA         | **62,00**         | **75,02**     | **86,80** | **0,00**   |
| Villanueva de los Infantes | **66,24**         | **71,48**     | **87,04** | 4,24       |
| Alcubillas                 | **55,88**         | **66,76**     | **86,64** | 8,26       |
| Torres de Montiel          | **66,36**         | **80,32**     | **95,68** | 8,88       |
| Cózar                      | **57,52**         | **77,72**     | **95,72** | 8,92       |
| Fuenllana                  | **71,56**         | **76,36**     | **87,00** | 9,56       |

Tabla 21. Distancias $d_\infty$ a la combinación en kilómetros correspondiente a (2,2.42,2.8) jornadas, para las cinco localidades con combinación más cercana.

|                            | Venta de Cárdenas | Puerto Lápice | El Toboso | $d_1$   |
|----------------------------|-------------------|---------------|-----------|---------|
|                            | **km**            | **km**        | **km**    |         |
| LUGAR DE LA MANCHA         | **62,00**         | **75,02**     | **86,80** | **0,00**|
| Villanueva de los Infantes | **66,24**         | **71,48**     | **87,04** | 8,02    |
| Fuenllana                  | **71,56**         | **76,36**     | **87,00** | 11,10   |
| Alcubillas                 | **55,88**         | **66,76**     | **86,64** | 14,54   |
| Cózar                      | **57,52**         | **77,72**     | **95,72** | 16,10   |
| Torres de Montiel          | **66,36**         | **80,32**     | **95,68** | 18,54   |

Tabla 22. Distancias $d_1$ a la combinación en kilómetros correspondiente a (2,2.42,2.8) jornadas, para las cinco localidades con combinación más cercana.

|                            | Venta de Cárdenas | Puerto Lápice | El Toboso | $d_2$   |
|----------------------------|-------------------|---------------|-----------|---------|
|                            | **km**            | **km**        | **km**    |         |
| LUGAR DE LA MANCHA         | **62,00**         | **75,02**     | **86,80** | **0,00**|
| Villanueva de los Infantes | **66,24**         | **71,48**     | **87,04** | 5,53    |
| Fuenllana                  | **71,56**         | **76,36**     | **87,00** | 9,66    |
| Alcubillas                 | **55,88**         | **66,76**     | **86,64** | 10,28   |
| Cózar                      | **57,52**         | **77,72**     | **95,72** | 10,34   |
| Torres de Montiel          | **66,36**         | **80,32**     | **95,68** | 11,22   |

Tabla 23. Distancias $d_2$ a la combinación en kilómetros correspondiente a (2,2.42,2.8) jornadas, para las cinco localidades con combinación más cercana.

## 6.4. Análisis temporal con 3 referencias

En esta subsección se realiza un estudio análogo al de la Subsección 6.2 pero eliminando en este caso los datos correspondientes a Munera. Las Tablas 24, 25 y 26 muestran los resultados de estas comparaciones utilizando las métricas $L_\infty$, $L_1$ y $L_2$ respectivamente.

|  | Venta de Cárdenas | Puerto Lápice | El Toboso | $d_\infty$ |
|---|---|---|---|---|
|  | **horas** | **horas** | **horas** |  |
| LUGAR DE LA MANCHA | **20,00** | **24,20** | **28,00** | **0,00** |
| Villanueva de los Infantes | **21,37** | **23,06** | **28,08** | 1,37 |
| Alcubillas | **18,03** | **21,54** | **27,95** | 2,66 |
| Torres de Montiel | **21,41** | **25,91** | **30,86** | 2,86 |
| Cózar | **18,55** | **25,07** | **30,88** | 2,88 |
| Fuenllana | **23,08** | **24,63** | **28,06** | 3,08 |

Tabla 24. Distancias $d_\infty$ a la combinación en horas correspondiente a (2,2.42,2.8) jornadas, para las cinco localidades con combinación más cercana.

|  | Venta de Cárdenas | Puerto Lápice | El Toboso | $d_1$ |
|---|---|---|---|---|
|  | **horas** | **horas** | **horas** |  |
| LUGAR DE LA MANCHA | **20,00** | **24,20** | **28,00** | **0,00** |
| Villanueva de los Infantes | **21,37** | **23,06** | **28,08** | 2,59 |
| Fuenllana | **23,08** | **24,63** | **28,06** | 3,57 |
| Alcubillas | **18,03** | **21,54** | **27,95** | 4,68 |
| Cózar | **18,55** | **25,07** | **30,88** | 5,20 |
| Torres de Montiel | **21,41** | **25,91** | **30,86** | 5,98 |

Tabla 25. Distancias $d_1$ a la combinación en horas correspondiente a (2,2.42,2.8) jornadas, para las cinco localidades con combinación más cercana.

|  | Venta de Cárdenas | Puerto Lápice | El Toboso | $d_2$ |
|---|---|---|---|---|
|  | **horas** | **horas** | **horas** |  |
| LUGAR DE LA MANCHA | **20,00** | **24,20** | **28,00** | **0,00** |
| Villanueva de los Infantes | **21,37** | **23,06** | **28,08** | 1,78 |
| Fuenllana | **23,08** | **24,63** | **28,06** | 3,11 |
| Alcubillas | **18,03** | **21,54** | **27,95** | 3,31 |
| Cózar | **18,55** | **25,07** | **30,88** | 3,34 |
| Torres de Montiel | **21,41** | **25,91** | **30,86** | 3,62 |

Tabla 26. Distancias $d_2$ a la combinación en horas correspondiente a (2,2.42,2.8) jornadas, para las cinco localidades con combinación más cercana.

**7. Recopilación de resultados y comparación con los trabajos previos**

Ahora recopilaremos los resultados obtenidos en este trabajo y los compararemos con los obtenidos en los trabajos previos. Para ello se va a utilizar la noción matemática de *error relativo*, cuya utilidad se comprende fácilmente si entendemos que no es lo mismo cometer un error de 100 gramos al pesar 1 kilogramo de una sustancia que cometerlo al pesar 100 kilogramos de esa misma sustancia. El error relativo en cada caso se define como 100/1000 y 100/100000, respectivamente, y consiste en relativizar el error absoluto (100 gramos) por la magnitud de lo medido (1000 ó 100000 gramos).

De manera similar, en la situación tratada en las secciones anteriores para cada localidad tenemos una distancia *d* a la combinación solución, que relativizaremos por la magnitud de esta combinación solución, calculada como su distancia a la combinación (0,0,0,0). Un tratamiento análogo se puede aplicar a los resultados obtenidos en los trabajos previos, como queda recogido en la Tabla 27.

|  | Localidad 1 |  | Localidad 2 |  | Localidad 3 |  |
|---|---|---|---|---|---|---|
| [7] | Alcubillas | 8,30 | Villanueva Inf. | 10,38 |  |  |
| [3] con $L_\infty$ | Fuenllana | 12,00 | Villanueva Inf. | 12,19 | Carrizosa | 15,12 |
| [3] con $L_1$ | Carrizosa | 6,86 | Fuenllana | 9,24 | Villanueva Inf. | 9,27 |
| [3] con $L_2$ | Carrizosa | 9,15 | Villanueva Inf. | 9,90 | Fuenllana | 9,98 |
| Subsecc. 5.1 ($L_\infty$) | Alcubillas | 11,79 | Carrizosa | 12,23 | Villanueva Inf. | 12,31 |
| Subsecc. 5.1 ($L_1$) | Villanueva Inf. | 6,10 | Carrizosa | 7,06 | Alcubillas | 9,84 |
| Subsecc. 5.1 ($L_2$) | Villanueva Inf. | 7,72 | Carrizosa | 9,23 | Alcubillas | 10,02 |
| Subsecc. 5.2 ($L_\infty$) | Alcubillas | 11,80 | Carrizosa | 12,24 | Villanueva Inf. | 12,32 |
| Subsecc. 5.2 ($L_1$) | Villanueva Inf. | 6,10 | Carrizosa | 7,06 | Alcubillas | 9,83 |
| Subsecc. 5.2 ($L_2$) | Villanueva Inf. | 7,74 | Carrizosa | 9,22 | Alcubillas | 10,03 |
| Subsecc. 5.3 ($L_\infty$) | Carrizosa | 10,89 | Alcubillas | 11,79 | Villanueva Inf. | 12,31 |
| Subsecc. 5.3 ($L_1$) | Carrizosa | 4,66 | Alcubillas | 7,40 | Villanueva Inf. | 10,31 |
| Subsecc. 5.3 ($L_2$) | Carrizosa | 6,91 | Villanueva Inf. | 8,61 | Alcubillas | 10,43 |
| Subsecc. 5.4 ($L_\infty$) | Carrizosa | 10,88 | Alcubillas | 11,80 | Villanueva Inf. | 12,32 |
| Subsecc. 5.4 ($L_1$) | Carrizosa | 4,66 | Villanueva Inf. | 7,41 | Fuenllana | 10,29 |
| Subsecc. 5.4 ($L_2$) | Carrizosa | 6,90 | Villanueva Inf. | 8,61 | Alcubillas | 10,42 |
| Subsecc. 6.1 ($L_\infty$) | Villanueva Inf. | 9,37 | Alcubillas | 9,52 | Cózar | 10,28 |
| Subsecc. 6.1 ($L_1$) | Villanueva Inf. | 5,51 | Alcubillas | 5,66 | Cózar | 6,07 |
| Subsecc. 6.1 ($L_2$) | Villanueva Inf. | 6,66 | Cózar | 7,10 | Alcubillas | 7,10 |
| Subsecc. 6.2 ($L_\infty$) | Villanueva Inf. | 9,36 | Alcubillas | 9,50 | Cózar | 10,29 |
| Subsecc. 6.2 ($L_1$) | Villanueva Inf. | 5,51 | Alcubillas | 5,65 | Cózar | 6,08 |
| Subsecc. 6.2 ($L_2$) | Villanueva Inf. | 6,66 | Alcubillas | 7,10 | Cózar | 7,10 |
| Subsecc. 6.3 ($L_\infty$) | Villanueva Inf. | 4,88 | Alcubillas | 9,52 | Torres de M. | 10,23 |
| Subsecc. 6.3 ($L_1$) | Villanueva Inf. | 3,58 | Fuenllana | 4,96 | Alcubillas | 6,50 |
| Subsecc. 6.3 ($L_2$) | Villanueva Inf. | 4,24 | Fuenllana | 7,41 | Alcubillas | 7,88 |
| Subsecc. 6.4 ($L_\infty$) | Villanueva Inf. | 4,89 | Alcubillas | 9,50 | Torres de M. | 10,21 |
| Subsecc. 6.4 ($L_1$) | Villanueva Inf. | 3,59 | Fuenllana | 4,94 | Alcubillas | 6,48 |
| Subsecc. 6.4 ($L_2$) | Villanueva Inf. | 4,23 | Fuenllana | 7,39 | Alcubillas | 7,87 |

Tabla 27. De izquierda a derecha, las tres localidades más cercanas a la solución junto con el error relativo en tanto por ciento. Se incluyen tanto los trabajos previos [7] y [3], como las Secciones 5 y 6 del presente trabajo, sombreadas en distinto color según utilicen kilómetros u horas.

En esta tabla se observa que los errores relativos más pequeños son los correspondientes a las Subsecciones 6.3 y 6.4 del presente trabajo, es decir, los correspondientes a no considerar Munera como punto de referencia y utilizar la combinación solución de (2,2.42,2.8,2.23) jornadas introducida en [6]. Estos son también los casos con las mayores diferencias entre el error relativo de la primera localidad candidata y el de la segunda, tanto en términos absolutos como al considerar la media de estas diferencias para las tres métricas; véase la Tabla 28.

|  | Localidad 2 – Localidad 1 | Media |
|---|---|---|
| [7] | 2,08 | |
| [3] con $L_\infty$ | 0,19 | |
| [3] con $L_1$ | 2,37 | 1,10 |
| [3] con $L_2$ | 0,75 | |
| Subsecc. 5.1 ($L_\infty$) | 0,44 | |
| Subsecc. 5.1 ($L_1$) | 0,96 | 0,97 |
| Subsecc. 5.1 ($L_2$) | 1,51 | |
| Subsecc. 5.2 ($L_\infty$) | 0,44 | |
| Subsecc. 5.2 ($L_1$) | 0,96 | 0,96 |
| Subsecc. 5.2 ($L_2$) | 1,48 | |
| Subsecc. 5.3 ($L_\infty$) | 0,90 | |
| Subsecc. 5.3 ($L_1$) | 2,74 | 1,78 |
| Subsecc. 5.3 ($L_2$) | 1,70 | |
| Subsecc. 5.4 ($L_\infty$) | 0,92 | |
| Subsecc. 5.4 ($L_1$) | 2,75 | 1,79 |
| Subsecc. 5.4 ($L_2$) | 1,71 | |
| Subsecc. 6.1 ($L_\infty$) | 0,15 | |
| Subsecc. 6.1 ($L_1$) | 0,15 | 0,25 |
| Subsecc. 6.1 ($L_2$) | 0,44 | |
| Subsecc. 6.2 ($L_\infty$) | 0,14 | |
| Subsecc. 6.2 ($L_1$) | 0,14 | 0,24 |
| Subsecc. 6.2 ($L_2$) | 0,44 | |
| Subsecc. 6.3 ($L_\infty$) | 4,63 | |
| Subsecc. 6.3 ($L_1$) | 1,38 | 3,06 |
| Subsecc. 6.3 ($L_2$) | 3,17 | |
| Subsecc. 6.4 ($L_\infty$) | 4,61 | |
| Subsecc. 6.4 ($L_1$) | 1,36 | 3,04 |
| Subsecc. 6.4 ($L_2$) | 3,16 | |

Tabla 28. Diferencia entre la segunda localidad candidata y la primera, junto con la media de estas diferencias para las tres métricas en cada una de las configuraciones.

## 8. Conclusiones

Tras reiterar nuestra intención de dejar para los expertos en otras disciplinas la discusión sobre la corrección de los datos aportados en [5], [6] y [8], en este trabajo nos hemos limitado a realizar un análisis matemático básico de los mismos. De este análisis se pueden extraer las siguientes conclusiones:

- Si se considera la combinación solución (2,2.37,2.5,2) comúnmente utilizada hasta el momento (Sección 5 de este trabajo), los datos de [8] arrojan resultados dispares cuando se incluye Munera como punto de referencia, obteniéndose como localidad candidata Villanueva de los Infantes para las métricas $L_1$ y $L_2$, y Alcubillas para la métrica $L_\infty$. Si bien esta métrica es menos representativa que las otras dos, incluso para éstas la segunda localidad candidata tiene un error relativo muy similar a la primera, lo que hace poco significativo su resultado.

- Si para la misma combinación solución se excluye Munera como punto de referencia, todas las métricas coinciden al determinar Carrizosa como localidad candidata. Las diferencias entre la primera localidad y la segunda en este caso son las segundas más significativas de la Tabla 28.
- Si se considera la combinación solución (2,2.42,2.8,2.23) introducida en [6] (Sección 6 de este trabajo), para todos los casos estudiados los datos de [8] establecen Villanueva de los Infantes como localidad candidata.
- En particular, Villanueva de los Infantes es la localidad candidata para la configuración con los errores relativos más pequeños y las mayores diferencias entre la primera localidad candidata y la segunda. Esta configuración es la que no considera Munera como punto de referencia y utiliza la combinación solución (2,2.42,2.8,2.23) introducida en [6] (Subsecciones 6.3 y 6.4 del presente trabajo).
- El hecho de no considerar Munera como punto de referencia mejora, siempre según los datos de [8], los errores relativos y las diferencias entre la primera localidad candidata y la segunda.
- El análisis de los datos de [8] expresados en kilómetros (Tabla 1, Subsecciones 5.1, 5.3, 6.1 y 6.3) ofrece resultados casi idénticos al uso de los datos de [8] expresados en horas (Tabla 5, Subsecciones 5.2, 5.4, 6.2 y 6.4).
- La combinación solución (2,2.42,2.8,2.23) introducida en [6] da lugar a los mejores resultados del presente trabajo, en términos de magnitud del error relativo y diferencia entre las dos primeras localidades candidatas, cuando no se considera Munera como punto de referencia (Subsecciones 6.3 y 6.4). Sin embargo, también da lugar a los peores resultados de este trabajo, cuando sí se considera esa referencia (Subsecciones 6.1 y 6.2).
- La Tabla 28 muestra que, para la mayoría de las configuraciones estudiadas en el presente trabajo, los datos de [8] dan lugar a resultados que mejoran los obtenidos en los trabajos previos, en términos de magnitud del error relativo y diferencia entre las dos primeras localidades candidatas.

## 9. Bibliografía.